\documentclass[preprint]{elsarticle}

\usepackage{amsmath} 
\usepackage{amsfonts}
\usepackage{graphicx}
\usepackage{color}
\usepackage{gensymb}

\usepackage{hyperref}

\hypersetup{
     unicode=false,          
     pdftoolbar=true,        
     pdfmenubar=true,        
     pdffitwindow=false,     
     pdfstartview={FitH},    
     pdfauthor={CAST Collaboration},     
     pdfsubject={PWA},       
     pdfnewwindow=true,      
     colorlinks=false,       
     linkcolor=red,          
     citecolor=green,        
     filecolor=magenta,      
     urlcolor=cyan           
     }

\journal{Physics of the Dark Universe}

\begin{document}

\begin{frontmatter}
%
%
\title{First Results on the Search for Chameleons with the KWISP Detector at CAST}
%
%
\author[cern]{ S. Arguedas Cuendis}
\author[freiburg]{    J.~Baier  } 
\author[cern]{    K.~Barth  } 
\author[klein,nordita]{    S.~Baum }
\author[bilgi,x]{A.~Bayirli}
\author[inr]{    A.~Belov  }  
\author[mpe]{    H.~Br\"auninger  } 
\author[trieste,triesteu]{    G.~Cantatore*  }  
\author[zaragoza]{    J.~M.~Carmona  } 
\author[zaragoza]{    J.~F.~Castel  }  
\author[bilgi]{    S.~A.~Cetin  } 
\author[zaragoza]{    T.~Dafni  }  
\author[cern]{    M.~Davenport  }  
\author[inr]{    A.~Dermenev  }  
\author[bonn]{	K.~Desch}  
\author[cern]{    B.~D\"obrich  } 
\author[freiburg]{    H.~Fischer*  }  
\author[cern]{    W.~Funk  }  
\author[zaragoza]{    J.~A.~Garc\' ia  }
\author[patras]{    A.~Gardikiotis  } 
\author[zaragoza]{    J.~G.~Garza  } 
\author[inr]{    S.~Gninenko  }  
\author[vancouver]{    M.~D.~Hasinoff  }  
\author[xian]{    D.~H.~H.~Hoffmann  }  
\author[zaragoza]{   F.~J.~Iguaz  }  
\author[zaragoza]{   I.~G.~Irastorza  }  
\author[zagreb]{   K.~Jakov\v ci\' c  }  
\author[bonn]{	J. Kaminski}  
\author[rijeka,cerijeka,trieste]{   M.~Karuza* } 
\author[bonn,§§]{ 	C. Krieger}  
\author[zagreb]{    B.~Laki\'{c}  }  
\author[cern]{    J.~M.~Laurent  }  
\author[zaragoza]{    G.~Luz\'on  }  
\author[patras]{    M.~Maroudas}  
\author[capp]{    L.~Miceli  }  
\author[darmstadt]{    S.~Neff  }  
\author[zaragoza,cern]{    I.~Ortega  } 
\author[bilgi,§]{A.~Ozbey}
\author[llnl]{    M.~J.~Pivovaroff  }  
\author[eli]{    M.~Rosu  }
\author[llnl]{    J.~Ruz  } 
\author[zaragoza]{    E.~Ruiz Ch\'oliz  }  
\author[bonn]{    S.~Schmidt  } 
\author[freiburg]{M. Schumann }
\author[capp,kaist]{    Y.~K.~Semertzidis  }
\author[mpis]{    S.~K.~Solanki  }
\author[cern]{    L.~Stewart  }  
\author[patras]{    I.~Tsagris} 
\author[cern]{    T.~Vafeiadis  } 
\author[llnl]{    J.~K.~Vogel  } 
\author[rijeka]{    M.~Vretenar  }  
\author[bilgi,∆]{    S.~C.~Yildiz  }
\author[patras,cern]{    K.~Zioutas  } 
%
\address[ihep]{Institute of High Energy Physics, Chinese Academy of Sciences, Beijing, China}
\address[bonn]{Physikalisches Institut, University of Bonn, 53115 Bonn, Germany.}
\address[capp]{Center for Axion and Precision Physics Research, Institute for Basic Science (IBS), Daejeon 34141, Republic of Korea.}
\address[kaist]{Department of Physics, Korea Advanced Institute of Science and Technology (KAIST), Daejeon 34141, Republic of Korea}
\address[freiburg]{Physikalisches Institut, Albert-Ludwigs-Universit\"{a}t Freiburg, 79104 Freiburg, Germany}
\address[mpe]{Max-Planck-Institut f\"{u}r Extraterrestrische Physik, Garching, Germany}
\address[cern]{European Organization for Nuclear Research (CERN), Gen\`eve, Switzerland}
\address[mpis]{Max-Planck-Institut f\"{u}r Sonnensystemforschung, G\"{o}ttingen, Germany}
\address[bilgi]{Istanbul Bilgi University, High Energy Physics Research Center, Eyupsultan, Istanbul, Turkey}
\address[llnl]{Lawrence Livermore National Laboratory, Livermore, CA 94550, USA}
\address[eli]{Extreme Light Infrastructure - Nuclear Physics (ELI-NP), 077125 Magurele, Romania}
\address[inr]{Institute for Nuclear Research (INR), Russian Academy of Sciences, Moscow, Russia}
\address[patras]{Physics Department, University of Patras, Patras, Greece}
\address[rijeka]{University of Rijeka, Department of Physics, Rijeka, Croatia.}
\address[cerijeka]{Photonics and Quantum Optics Unit, Center of Excellence for Advanced Materials and Sensing Devices, and Centre for Micro and Nano Sciences and Technologies, University of Rijeka, Rijeka, Croatia }
\address[klein]{The Oskar Klein Centre for Cosmoparticle Physics, Department of Physics, Stockholm University, Alba Nova, 10691 Stockholm, Sweden}
\address[nordita]{Nordita, KTH Royal Institute of Technology and Stockholm University, Roslagstullsbacken 23, 10691 Stockholm, Sweden}
\address[trieste]{Istituto Nazionale di Fisica Nucleare (INFN), Sezione di Trieste, Trieste, Italy} 
\address[triesteu]{Universit\`a di Trieste, Trieste, Italy}
\address[vancouver]{Department of Physics and Astronomy, University of British Columbia, Vancouver, Canada}
\address[zagreb]{Rudjer Bo\v{s}kovi\'{c} Institute, Zagreb, Croatia}
\address[zaragoza]{Grupo de Investigaci\'{o}n de F\'{\i}sica Nuclear y Astropart\'{\i}culas, Universidad de Zaragoza,Zaragoza, Spain }
\address[xian]{Xi'An Jiaotong University, School of Science, Xi'An, 710049, China}
%
%
\fntext[x]{Present address: Bogazici University, Physics Department, Bebek, Istanbul, Turkey}
\fntext[§]{Present address: Istanbul University-Cerrahpasa, Voc. Sch. of Technical Sciences, Buyukcekmece, Istanbul, Turkey}
\fntext[∆]{Present address: Department of Physics and Astronomy, University of California, Irvine, California 92697, USA}
\fntext[§§]{Present address: Institute of Experimental Physics, University of Hamburg, 22761 Hamburg, Germany}
%
%
%
%
\cortext[mycorrespondingauthor]{Corresponding authors: \\ Giovanni.Cantatore@cern.ch, Horst.Fischer@cern.ch, Marin.Karuza@cern.ch}
%
%
\begin{abstract}
We report on a first measurement with a sensitive opto-mechanical force sensor designed for the direct detection of coupling of real chameleons to matter. These dark energy candidates could be produced in the Sun and stream unimpeded to Earth. 
The KWISP detector installed on the CAST axion search experiment at CERN looks for tiny displacements of a thin membrane caused by the mechanical effect of solar chameleons. The displacements are detected by a Michelson interferometer with a homodyne readout scheme.
The sensor benefits from the focusing action of the ABRIXAS X-ray telescope installed at CAST, which increases the chameleon flux on the membrane. A mechanical chopper placed between the telescope output and the detector modulates the incoming chameleon stream. 
We present the results of the solar chameleon measurements taken at CAST in July 2017, setting an upper bound on the force acting on the membrane of $80$~pN at 95\% confidence level. 
The detector is sensitive for direct coupling to matter $10^4 \leq\beta_m \leq 10^8$, where the coupling to photons is locally bound to $\beta_\gamma \leq 10^{11}$.
\end{abstract}
\end{frontmatter}

\pagebreak

\setcounter{page}{1}


\section{Introduction}
Chameleons are Weakly Interacting Slim Particles (WISPs) that, due to their scalar nature and density-dependent effective mass, are viable dark energy candidates \cite{art5}. Chameleons can couple both to photons, in analogy with the Sikivie coupling of axions \cite{Sikivie:1983jw}, and directly to matter, with the special property that their effective mass is dependent on the surrounding matter density. This allows the chameleon field to evade constraints set by "fifth-force" laboratory measurements \cite{fifthforce,fifthforce1}. Note that the first coupling requires the presence of a magnetic field, while the second, direct coupling to matter, does not. 

The following effective potential is used for chameleon description \cite{baum}:
\begin{equation}
	V_{eff}(\phi)=\Lambda^4 \left(1+\frac{\Lambda^n}{\phi^n}\right)+\rho_m e^{\frac{\beta_m \phi}{M_{Pl}}}+\frac{1}{4}F_{\mu\nu}F^{\mu\nu}e^{\frac{\beta_\gamma \phi}{M_{Pl}}} \label{eq:0}
\end{equation}
where $\Lambda$ is an energy scale, $\beta_m$ the coupling constant to matter, $M_{Pl}=2.435\cdot10^{18}$ GeV the reduced Planck mass, $\beta_\gamma$ the coupling constant to photons, $\rho_m$ the matter density surrounding the field and $F_{\mu\nu}$ the EM field strength. The last two terms in (\ref{eq:0}) represent the screening mechanism, the former describes coupling to matter, while the latter represents the coupling to photons.

The reflection \cite{reflection} of chameleon waves off a boundary between media of different densities at normal incidence has been extensively studied and can be naturally extended to the general case of incidence at an arbitrary angle \cite{art3}. The reflection of chameleons can be qualitatively explained in a simple way. In the low density region the effective mass of the chameleon is close to zero, and the main contribution to its energy comes from the momentum. If the chameleon energy is lower than the effective mass in the high density region, due to the energy conservation forbids chameleons to propagate in the dense region. Hence, incoming chameleons "bounce" off the boundary.

The effective chameleon mass $m$\ in a material of density $\rho_m$  is given by 

\begin{equation}
m=\left(n\left(n+1\right) \frac{\Lambda^{n+4}}{ \phi_{min}^{n+2}} \right)^{1/2}.
\label{eq:0.1}
\end{equation}

Here $\phi_{min}$ is the value of the scalar field for which the effective potential has a minimum, and is given by 
\begin{equation}
\phi_{min}=\left(\frac{n \Lambda^{n+4} M_{Pl}}{ \rho_m \beta_m} \right)^{1/{(n+1)}}.
\label{eq:0.1a}
\end{equation}

Note, reflected off a boundary surface chameleons deposit momentum, resulting in a net force being exerted on the surface itself \cite{baum}. It is this effect that the KWISP \cite{art4} (Kinetic WISP) force sensor attempts to exploit by searching for the force exerted on a thin $\mathrm{Si_3N_4}$ membrane by chameleons produced in the Sun. Tiny membrane displacements in response to applied forces are sensed with optical interferometry, in this particular case using a Michelson-type interferometer with a balanced homodyne readout. Here the membrane acts as a mirror in one of the arms of the interferometer.

The reflection properties of chameleons are also exploited in the design of a "chameleon chopper" that is used to modulate the amplitude of a chameleon beam at a given frequency. The modulated beam in turn generates a periodic force when reflecting off the $\mathrm{Si_3N_4}$ membrane, which can then be sensed in a frequency bin known \textit{a priori}, therefore enhancing the sensitivity of the experiment. In essence, the "chameleon chopper" is a device that periodically intercepts the chameleon beam with a material surface at a chosen angle which is selected to be compatible with the detector sensitivity to chameleons. The chameleon flux is thereby reduced, but when the surface is removed from the beam it again assumes the maximum value. In the setup described below, chopping is achieved by rapidly turning a wheel presenting alternatively reflecting and non-reflecting sectors to the beam.

To further improve the chances of detection, the ABRIXAS X-ray telescope \cite{art7}, available at CAST, is exploited to increase the chameleon flux on the membrane by a factor of the order of 100, in our case, where only the CAST magnet bore is considered as the viable chameleon path. The KWISP membrane is positioned in the focal plane of the telescope in order to take advantage of its focusing action, while the chameleon chopper is set between the telescope output and the sensing membrane.
The signature of a particle flux from the Sun would be the observation of a signal at the chopper modulation frequency during periods when the telescope is aligned with the Sun, combined with non-observation during background measurements, when the telescope points away from the Sun.

In the following, we expand on the technical details of our detector setup and of the chopper, discuss the sensor characterisation done in an optics laboratory, and present the results of the solar tracking measurements carried out at CAST at the end of June, beginning of July 2017.

\section{Setup} 
The KWISP detector presented in this paper consists of a rigid,  100~nm thick, dielectric, 5x5 $\mathrm{mm^2}$ $\mathrm{Si_3N_4}$ membrane (made by Norcada Inc., Canada, part no. QX10500CS) mounted as a mirror in one arm of a Michelson interferometer. The arm length is approximately 7 cm. Membrane movements cause a change in the Optical Path Length (OPL) difference between the two arms of the Michelson interferometer. This difference in OPL translates into a phase shift appearing as a change in the intensity of the fringes at the interferometer output. The intensity of the fringes is read out using a balanced homodyne detection scheme. This allows for sensitive detection of small displacements of the membrane and thus the detection of tiny forces exerted on the membrane. 
The homodyne readout scheme was chosen also because of its intrinsic 60 dB common-noise rejection ratio, allowing the use of different light sources independently of their noise characteristics.

\begin{figure}[ht!]
\centering
\includegraphics[width=0.8\textwidth]{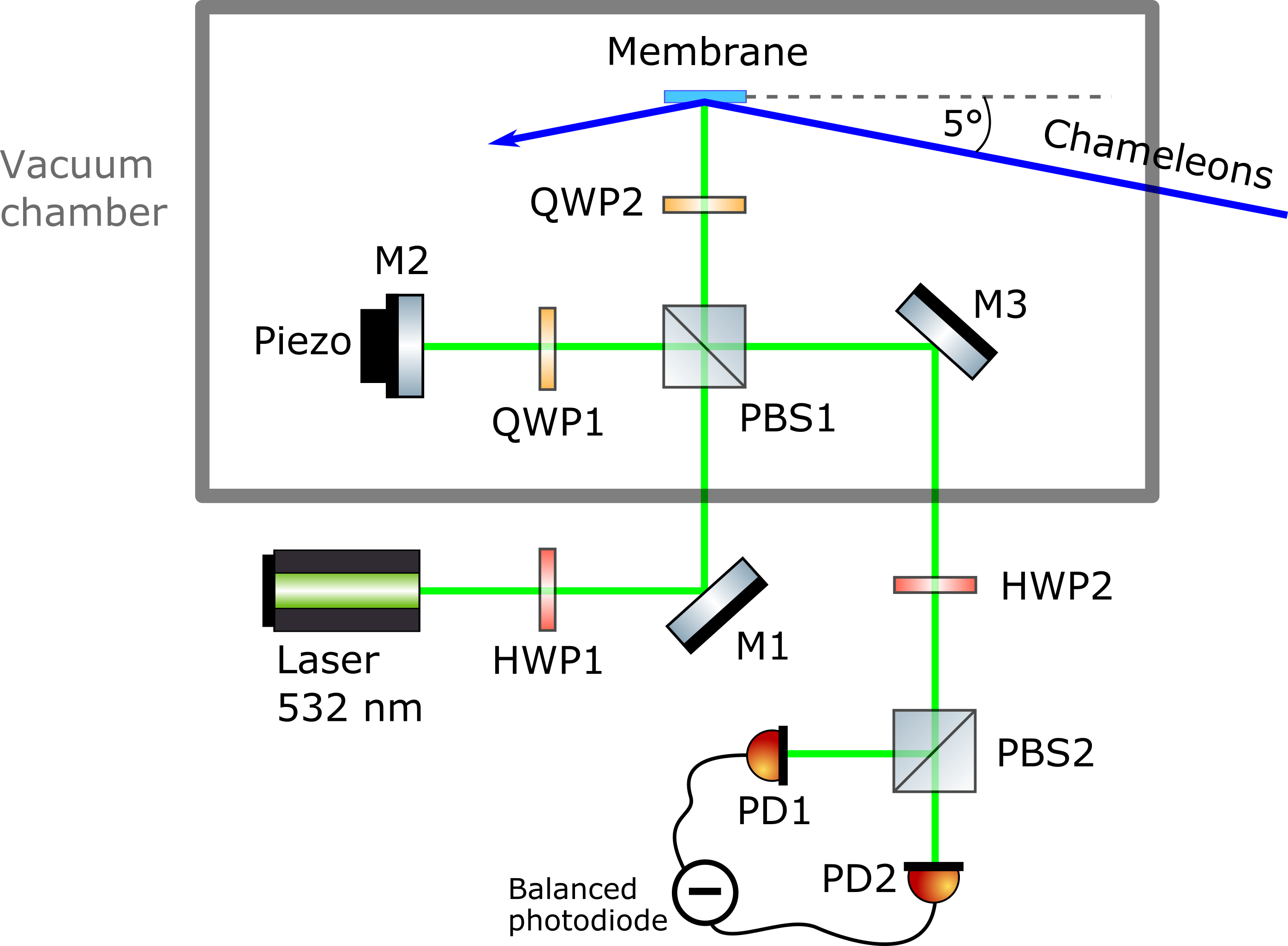}
\caption{Schematic layout of the the optical setup (see text).}
\label{pic:sch}
\end{figure}

The KWISP detector optical setup schematic is shown in Fig.~\ref{pic:sch}. A similar layout is also used in other high sensitivity measurements \cite{homo}. The laser beam is first sent through a half wave plate HWP1 which sets the ratio of the intensities of the beams in the two arms. The beam in one arm (\textit{uk - unknown beam}) goes to the membrane, while the other (\textit{lo - local oscillator}) goes to the mirror M2 connected to a piezo-electric actuator (PZT) which is used to phase lock this beam to the input beam. Each beam crosses twice a quarter-wave plate set at $45\degree$ changing the initial linear polarisation into a circular one, and after reflection on the respective mirrors and return path, back to a linear one, but rotated by $90\degree$ with respect to the original polarisation. This causes the returning beams to exit the polarising beam-splitter PBS1 at the other port. After recombination, the beams pass through a second half-wave plate HWP2 and a polarised beam splitter PBS2 to be evenly split and directed towards the two identical photo-detectors of the balanced detection scheme. The two signals output by the photo-detectors are electronically subtracted,  amplified  and then sent to the Data AcQuisition (DAQ). The pre-processing is entirely performed by the balanced detection photodiode (model PDB210A/M made by Thorlabs).

We now proceed to derive an expression for the signal output by the photo-detector. Assuming that the intensities of the local oscillator and of the unknown beam are the same before reentering PBS1, we may write their complex amplitudes as follows:
\begin{align}
\alpha_{lo}(t)&= \left[  \left(\frac{1}{\sqrt{2}}\,\alpha_0 + \delta X1_{lo}(t)\right) + i\delta X2_{lo}(t)\right]e^{i\left[\frac{2\pi l_{lo}}{\lambda}+\phi_{lo} \right]} \label{eq:1.1}\\
\alpha_{uk}(t)&= \left[ \left(\frac{1}{\sqrt{2}}\, \alpha_0 + \delta X1_{uk}(t)\right) + i\delta X2_{uk}(t)\right]e^{i\left[\frac{2\pi l_{uk}}{\lambda}+\phi_{m}(t) \right]}, \label{eq:1.2}
\end{align}
where $\alpha_0$ is the input beam amplitude, $\lambda$ its wavelength, and the variables $\delta X1_{lo}(t)$, $i\delta X2_{lo}(t)$, $\delta X1_{uk}(t)$ and $i\delta X2_{uk}(t)$ represent the amplitude and phase noise quadrature of the local oscillator and of the unknown beam, respectively. The quantities $l_{lo}$ and $l_{uk}$ represent the optical path lengths in the two interferometer arms. A membrane movement introduces a small phase shift $\phi_{m}(t)$ in the unknown beam. 

Since the phase shift introduced by the membrane is small, $\phi_{m}(t)\ll 1$, the following approximation may be used:
\begin{equation}
e^{i\phi_{m}(t)}=\cos\left[\phi_{m}(t)\right]+i\sin\left[\phi_{m}(t)\right]\approx 1+i\phi_{m}(t). \label{eq:1.3}
\end{equation}
Inserting Eq. (\ref{eq:1.3}) into Eq. (\ref{eq:1.2}) and neglecting terms containing $\phi_{m} \delta X $ gives:
\begin{equation}
\alpha_{uk}(t)= \left[  \left(\frac{1}{\sqrt{2}} \, \alpha_0+ \delta X1_{uk}(t)\right) + i\left(\delta X2_{uk}(t)+\frac{1}{\sqrt{2}} \,\alpha_0\,\phi_{m}(t)\right)\right]e^{i\frac{2\pi l_{uk}}{\lambda}}.  \label{eq:1.35}
\end{equation}
The two beams are recombined after PBS1 and then again split after PBS2 into two beams of equal intensities (balanced homodyne) having the complex amplitudes:
\begin{equation}
\alpha_{2,1}(t)=\frac{1}{\sqrt{2}}\alpha_{lo}(t)\pm\frac{1}{\sqrt{2}}\alpha_{uk}(t) \label{eq:1.4}
\end{equation}
The intensities of these beams are:
\begin{equation}
I_{2,1}(t)\propto |\alpha_{2,1}(t)|^2. \label{eq:1.5}
\end{equation}
Inserting Eqs. (\ref{eq:1.1}) and (\ref{eq:1.35}) into (\ref{eq:1.4}) and (\ref{eq:1.5}) and neglecting terms containing $\delta X\delta X$, we finally obtain the following expression for the difference photocurrent:
\begin{align}
i_-(t)\propto &\,I_2(t)-I_1(t) \nonumber\\
\propto &\left[ \alpha_0^2+\frac{2}{\sqrt{2}}\,\alpha_0\,\delta X1_{uk}(t) +\frac{2}{\sqrt{2}}\,\alpha_0\,\delta X1_{lo}(t) \right]\cos\left[\frac{2\pi}{\lambda}(l_{lo}-l_{uk})+\phi_{lo})\right]\nonumber\\
&+\left[ \alpha_0^2 \,\phi_m(t) +\frac{2}{\sqrt{2}}\,\alpha_0\,\delta X2_{uk}(t) +\frac{2}{\sqrt{2}}\,\alpha_0\,\delta X2_{lo}(t) \right]\sin\left[\frac{2\pi}{\lambda}(l_{lo}-l_{uk})+\phi_{lo})\right] .
\end{align}
The interferometer is held at a grey fringe by a piezoelectric actuator PZT controlled by a Proportional-Integral-Derivative (PID) feedback loop, that is, the following condition is maintained:
\begin{equation}
\frac{2\pi}{\lambda} (l_{lo}-l_{uk})+\phi_{lo}=\frac{\pi}{2} ,\label{eq:1.6}
\end{equation}
giving the resulting expression for the current signal at the output of the balanced detection photodiode amplifier,
\begin{equation}
i_-(t)\propto \alpha_0^2 \,\phi_m(t) + \frac{2}{\sqrt{2}}\alpha_0 \left[ \delta X2_{lo}(t)+ \delta X2_{uk}(t)\right],\label{eq:1}
\end{equation}
from which the membrane phase shift $\phi_m(t)$ may be obtained.

Different intensities and $I_{lo} / I_{uk}$ ratios were used during tests, as decreasing the ratio degrades sensitivity but lowers the influence of the light beam on the membrane. In the limit $I_{lo} \gg I_{uk}$, the expression of Eq. \ref{eq:1} becomes:
\begin{equation}
i_-(t)\propto 2\rho \alpha_0 \left[ \delta X2_{uk}(t)+\tau\,\alpha_0\,\phi_m(t) \right],\label{eq:2}.
\end{equation}
The parameters $\rho \approx 1$ and $\tau \ll 1$ are the normalized intensities of the local oscillator and of the unknown beams. With these settings the perturbation to the membrane due to the light radiation pressure is minimal. This was the case in the analysed run, where the dependence on the phase quadrature of the local oscillator is eliminated, and the limiting factor of the measurement becomes the shot noise of the unknown beam. The $I_{lo} \gg I_{uk}$ configuration is normally used to measure the quadratures of the unknown beam, for example in squeezing experiments \cite[and references therein]{bachor2004guide}.

The part of the optical setup placed inside a vacuum chamber (the interferometer, see Fig.~\ref{pic:sch}) is fixed on a single piece of aluminium designed for this purpose and the mirror movement needed to keep the optical path length difference always constant is provided by a PZT, glued directly between the mirror and its support.  The vacuum chamber is aligned with the CAST telescope in such a way that the chameleon stream passing through the chopper impinges on the membrane at an angle of $5\degree$.

A Field Programmable Gate Array (FPGA) development board called Red Pitaya \cite{RPTY} was used for the DAQ and setup control. It was chosen for its small size, which made it suitable for mounting directly on the CAST telescope, and for its local area network connectivity and analog-to-digital/digital-to-analog converter (ADC/DAC) capabilities.

\begin{figure}[ht!]
\centering
\includegraphics[width=1\textwidth]{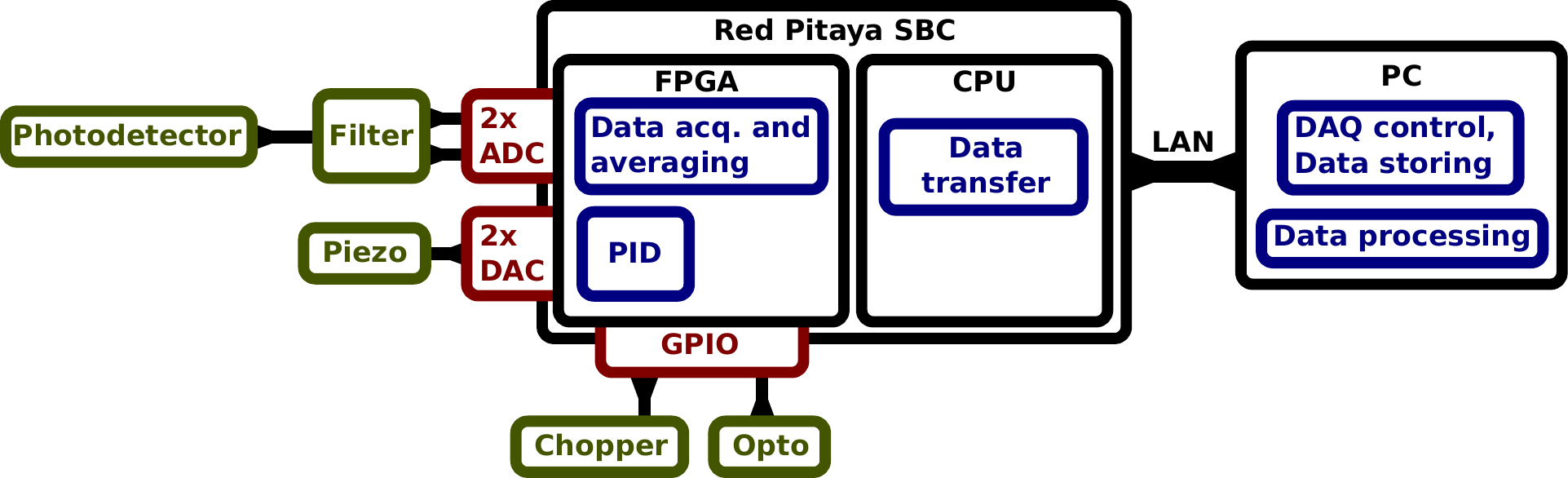}
\caption{DAQ block schematic.}
\label{pic:daq}
\end{figure}

Custom FPGA Verilog and C++ code was developed for it in order to enable continuous and lossless acquisition of data, along with an easy to use graphical user interface (GUI) program that allows control of all acquisition parameters and local storage of acquired data. The FPGA board also controls the PID feedback loop and the chopper frequency, additionally measuring it via a digital input connected to an optical switch. A block diagram of the DAQ is shown in Fig.~\ref{pic:daq}. The data are acquired at a 125 MS/s sample rate and adjacent data points are averaged in order to lower the waveform record length. This data stream is then transferred to a PC in the control room via LAN for on site analysis and storage. The raw data  also go in parallel to a PID controller whose output controls the mirror PZT, thus keeping the OPL difference at predetermined set point. The maximum unlimited continuous lossless acquisition  rate is $(125\; {\mathrm {MS/s}})/ 1024 \approx 122\; {\mathrm {kS/s}}$\ on both 14-bit ADC channels simultaneously, plus 1 bit chopper input which is well above any achievable mechanical chopper frequency. The ADC inputs have two modes, with the input range being $ \pm1 \ \mathrm{V_{pp}}$ and $\pm10 \ \mathrm{V_{pp}}$. However, both give the same signal-to-noise ratio confirming that we are not limited by the ADC resolution.

An additional GUI program was developed for performing fast Fourier transforms on acquired datasets. The results can be exported and displayed in various ways, also while overlaying the independently measured chopper frequency over the signal spectrum for quick online analysis and diagnostics.

\section{Solar chameleon flux}

\begin{figure}[ht!]
\centering
\includegraphics[width=1\textwidth]{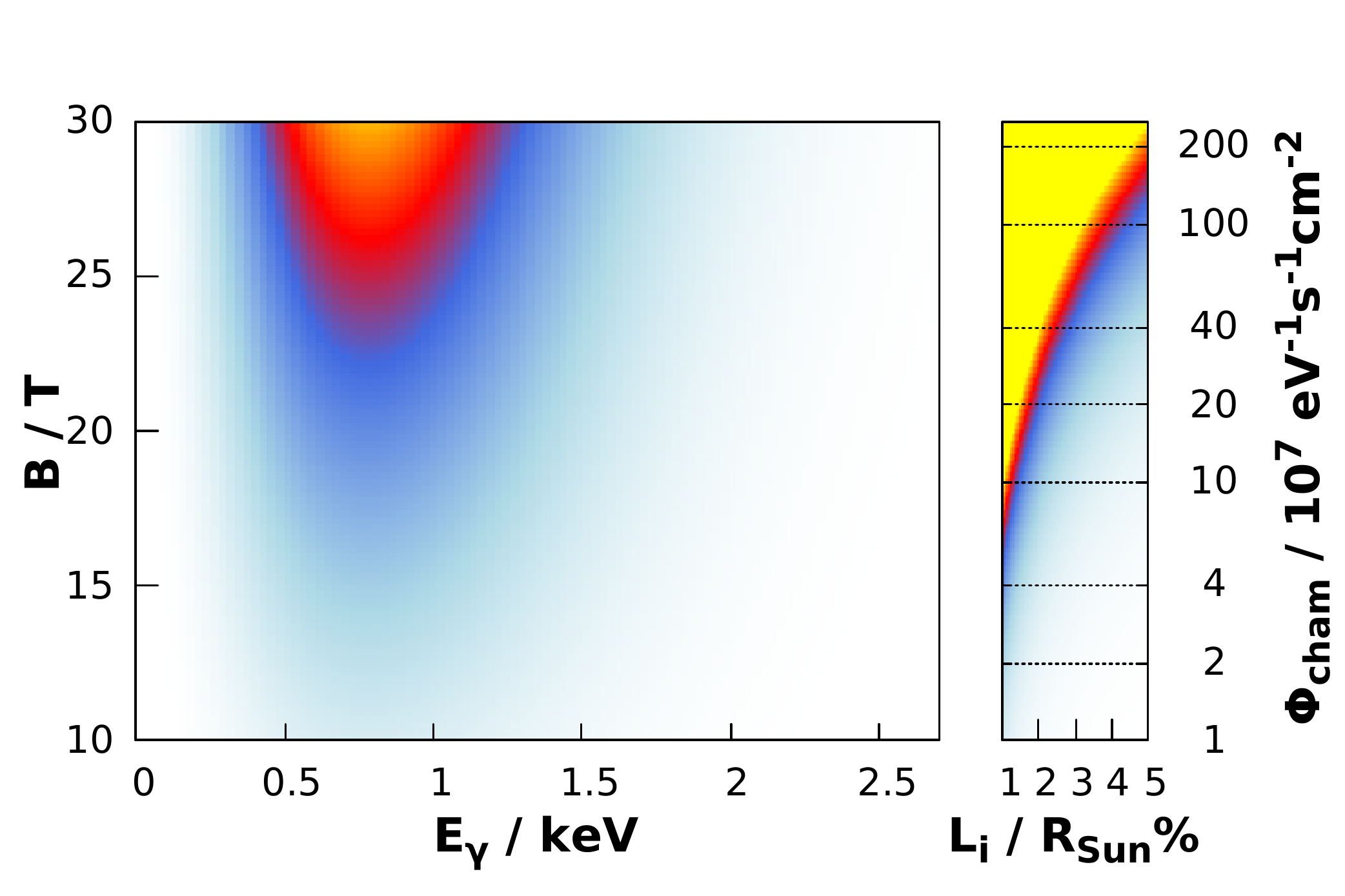}
\caption{A calculation of the chameleon flux at Earth. The interaction region length is taken in its uncertainty interval from 1\% to 5\% of solar radius. For a given pair of energy and field values in the left plot, the corresponding colour in the right plot provides the range of fluxes for various interaction region lengths.}
\label{pic:flux}
\end{figure}

In this and in the following sections we will present the essential steps in the evaluation of the expected chameleon flux reaching the KWISP detector. We start with the determination of the solar chameleon flux. Here we will give the main elements of the calculation, while more details can be found in separate articles\cite{brax, solar_cham}. The chameleon flux is a function of the Planckian distribution of thermal photons in the Sun $p_\gamma(\omega)=\frac{\omega^2}{\pi^2}\frac{1}{\kappa e^{\frac{\omega}{T}}-1}$, where $\kappa = \frac{2\zeta(3)T^3}{\pi^2}$, of the photon flux going through the magnetic region near the tachocline $n_\gamma$, and of the probability of creating chameleons out of thermal photons $P_{total}(\omega)$.
We assume that photon-chameleon conversion takes place in the tachocline region where the magnetic field strength in the sun is strongest, setting it to zero everywhere else. We use two values for the magnetic field, but provide a range of calculations in Fig.\ref{pic:flux}. 

Chameleons produced from photons with energy $\omega$ can only propagate in the Sun if their effective momentum $k$ satisfies
\begin{equation}
   k^2 = \omega^2 - \left( m^2 - \omega_{\rm pl}^2 \right) \geq 0\;.
\end{equation}
Here,  $m^2$ is the effective mass of the chameleon, cf. Eq.~\ref{eq:0.1}, and $\omega_{\rm pl}$ the plasma mass of the photon. The chameleon flux is then given by
\begin{equation}
\Phi_{cham}(\omega)=p_\gamma(\omega)P_{total}(\omega)n_\gamma\Theta(\omega^2-m^2-\omega_{pl}^2).
\label{eq:flux}
\end{equation}
Furthermore, the total probability $P_{total}(\omega)$ is proportional to the number of interactions inside the solar magnetic field $N=(\frac{L_i}{\lambda})^2$  and the square of the mixing angle $\theta$:
\begin{equation}
tan~ \theta = \left[\frac{B\omega\beta_\gamma} {M_{Pl}\left(m^2-(\frac{B\beta_\gamma}{M_{Pl}})^2-\omega^2_{pl}\right)} \right].
\end{equation}
\\
Here $\omega_{pl}$ is the plasma frequency and $\beta_\gamma$ is the coupling constant of chameleons to photons. The number of interactions depends on the length of the interaction region $L_i$, a thin shell near the tachocline of length $r \leq 0.05 R_{Sun}$, and on the inverse of the photon mean free path $\lambda = 0.25$~cm \cite{ppath}. The other quantities entering the evaluation of the chameleon flux are the number of photons crossing the tachocline  $n_\gamma = 3\cdot 10^{20} \frac{\mathrm{photons}}{\mathrm{s \ cm^2}}$ \cite{vinyoles}, the temperature of the tachocline $T = 2.3 \cdot 10^6 $~K, the magnetic field in the interaction region, taken either as $B=10$~T or as $B=30$~T , and the density of the Sun at the tachocline $\rho_{tacho} = 0.2 \frac{\mathrm{g}}{\mathrm{cm^3}}$. This expression is simplified by approximating $tan \theta \approx \theta$ since $\theta$ typically assumes values close to zero. Furthermore the chameleon model we considered assumes $n=1$ and the energy scale $\Lambda = 2.4 \cdot 10^{-3}$~eV. In all calculations the reduced Planck mass was used. An example of solar chameleons flux assuming $\beta_m=10^6$ is shown in Fig. \ref{pic:flux}. It can be verified that the total integrated flux satisfies the solar luminosity constraints.

\section{Propagation of the chameleons to the detector}

\begin{figure}[ht!]
\centering
\includegraphics[width=1\textwidth]{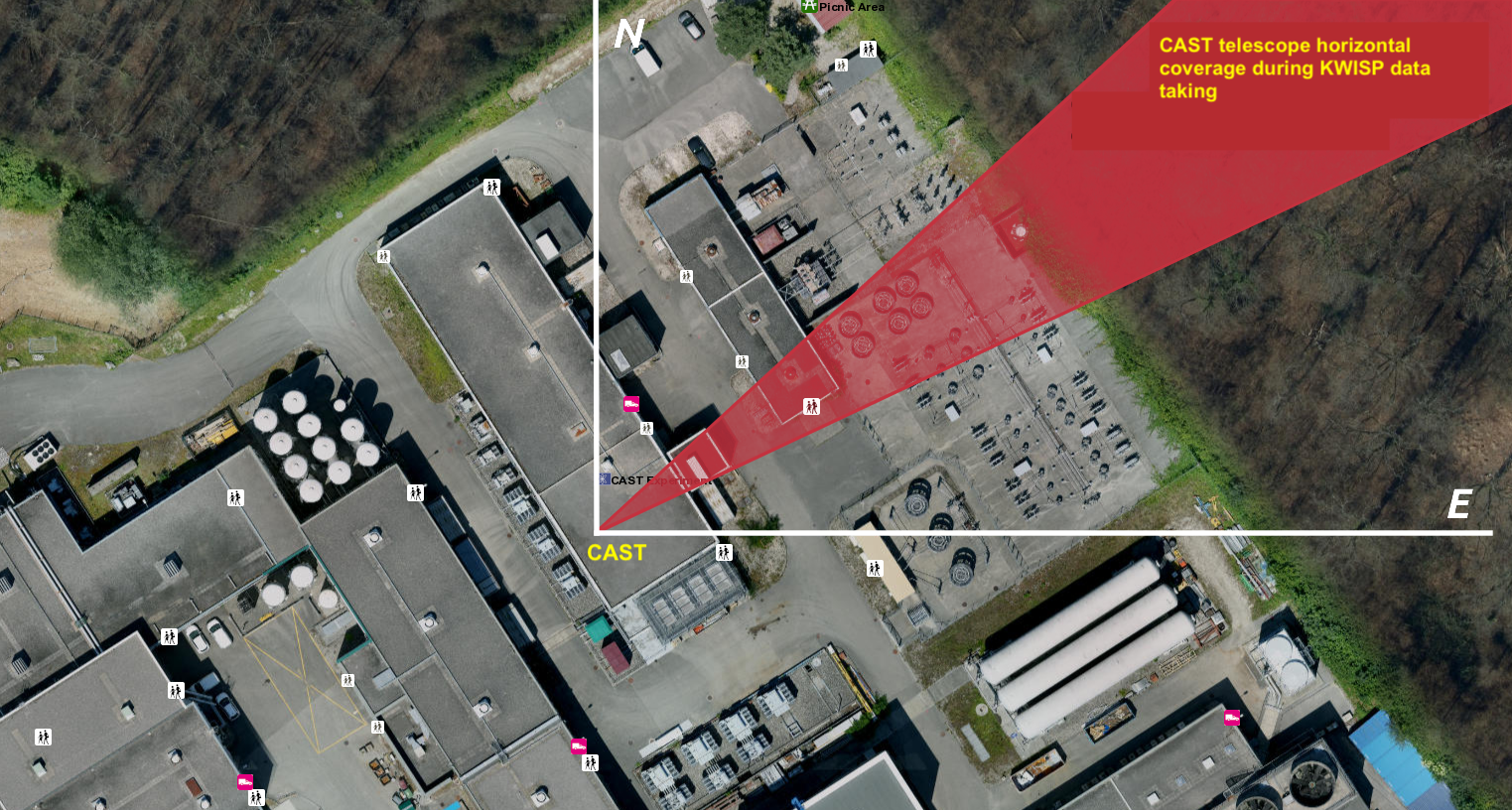}
\caption{Azimuthal coverage of the telescope on 3rd July 2017 during data taking. Picture of the CAST site courtesy of CERN.}
\label{pic:sun}
\end{figure}

The propagation of chameleons to Earth has been described previously \cite{baum}, but for completeness we will briefly go through the argument also in this work. The main obstacles to the chameleon propagation come either from a dense material on their path, where their effective mass could be higher than their energy, or a material with a smooth surface presented at a grazing incidence angle to the chameleon flux.
Once the chameleons leave the Sun they propagate freely to the Earth's atmosphere where they encounter for the first time a region of space with different density. If we model the atmosphere as a solid sphere of density $\rho=1.2\; \frac{\mathrm{kg}}{\mathrm{m^3}}$ with an average radius of the troposphere $R_{troposphere} = 18\; \mathrm{km}$, where 80\% of its mass is located, then the angle of incidence of a solar chameleon beam on the troposphere elongates at most $12\degree$ from normal incidence, this limit being dictated by the maximum $8\degree$ elevation angle of the CAST magnet. One can then safely assume that a solar chameleon beam always hits the troposphere at normal or close to normal incidence. After entering the atmosphere the chameleons do not encounter further density boundaries until reaching the surface of the Earth. Here we have two different cases. One case concerns telescope elevations above the horizon, and the other the telescope pointing below the horizon. In the former case the satellite picture shown in Fig. \ref{pic:sun}, combined with the position information from the CAST sun tracking control, shows that during data taking the only structures in front of the telescope are the walls of the experimental hall and of the adjacent buildings. These structures have not been studied in detail, however it is highly unlikely that they contain uniform surfaces of optical quality, which is a necessary condition to deflect the chameleon beam. Furthermore, these surfaces should be at grazing angle incidence, which is not the present case, at least when large structures are considered. 

The shading effect of the Earth for telescope elevations from $- 8\degree$ to $0\degree$ has not been previously discussed and was not included in past works \citep{castchin,ANASTASSOPOULOS2015172}. However, we will give here an argument to justify why in the zero order approximation this effect can be neglected. If we consider the Earth as a solid sphere, then the average incidence angle of the solar chameleon flux varies from $82\degree$ to $90\degree$ at the point where it encounters the surface of the Earth. Thus the main effect of the Earth at the chameleon flux is a reduction of intensity due to Earth's relatively high density. The highest density value the chameleons might encounter during the propagation through the Earth is that of peridotite, $\rho_p=3400\; \mathrm{kg/m^3}$. This is considerably lower than the density of steel $\rho_s=8100\; \mathrm{kg/m^3}$, which we have considered as the main source of shading effects. The propagation through the Earth's crust is not easy to simulate, and due to the fact that its interior is not uniform, the chameleons might encounter surfaces at different angles where diffusion can take place, however on average we expect that the direction and the intensity of the flux will remain constant. The energy loss during propagation through the Earth can be neglected since the recoiling mass of the Earth in this case is much larger than the chameleon mass. Therefore, the data for both under and over the horizon telescope elevations are indistinctly analysed.

\section{The chopper}
A constant pressure applied on a membrane simply sets it in a new static position, which cannot be distinguished from the non perturbed one since the difference in the DC power at the interferometer output is almost impossible to observe. On the other hand, a time-varying displacement generates a signal in a frequency region far from DC, which can be selected in a range where the measured background noise level is lower. Therefore, the chameleon stream must be modulated in amplitude before reaching the membrane: a periodically modulated chameleon flux exerts a periodic force on the membrane which causes it to vibrate at the modulation frequency. 

\begin{figure}[ht!]
\centering
\includegraphics[width=1\textwidth]{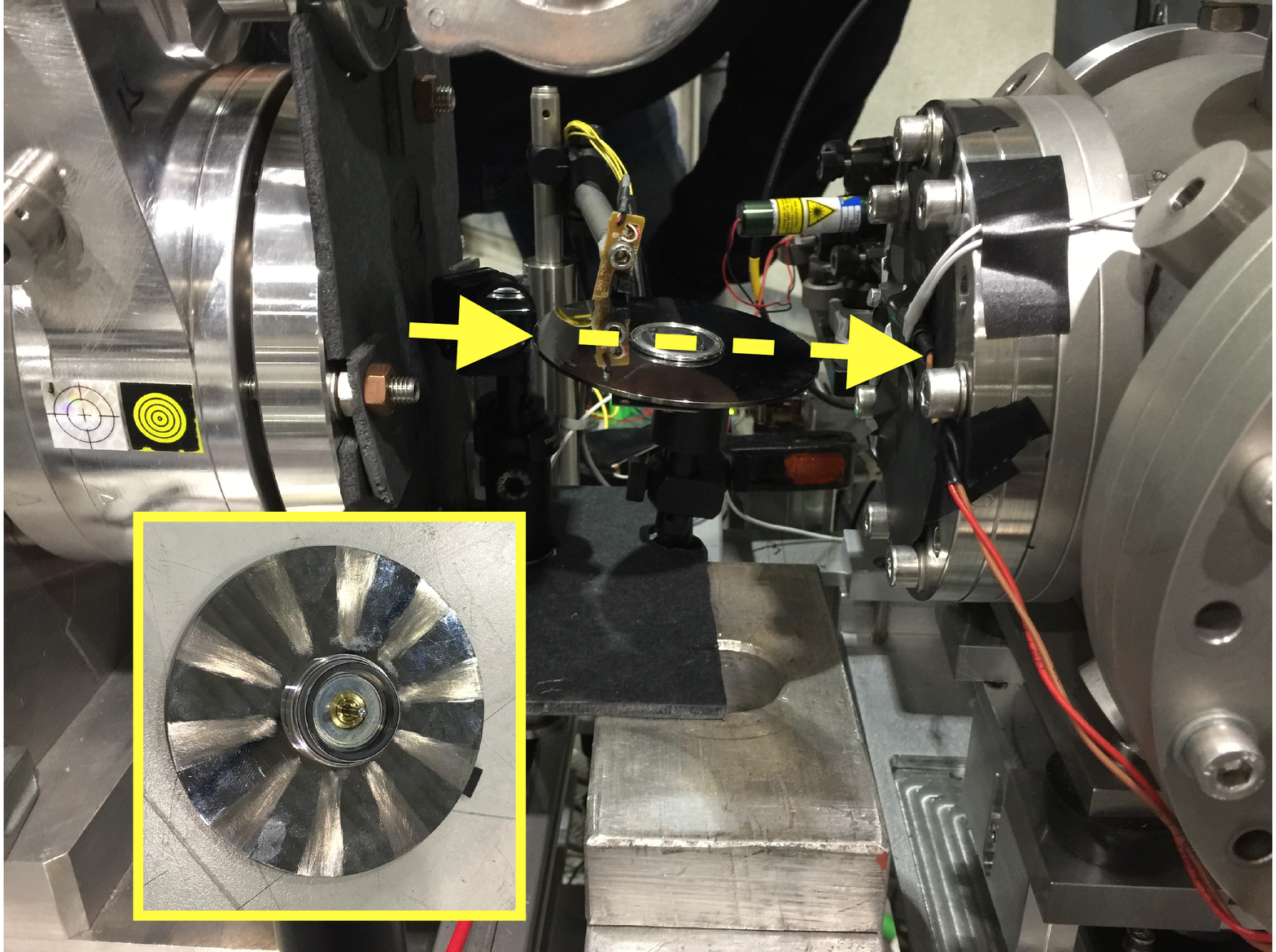}
\caption{The chopper disk mounted between the vacuum chamber (at right) and the telescope (at left). In the photograph the chameleon beam comes from the left hand side, as indicated by the arrows. The segmented surface of the disk faces downwards and it is illustrated in the framed inset.}
\label{pic:cho}
\end{figure}

For the measurements presented here, chameleon flux modulation is accomplished by exploiting the reflection of chameleons at grazing angles from a smooth metal surface. A rotating metal disk with alternating smooth and roughened sectors intercepts the chameleon stream at a grazing angle. The beam gains a periodic amplitude modulation, as it is expected to mostly scatter off the smooth or pass through the roughened areas, 
thus appearing turned alternatively off and on as seen by the membrane. A 3.5 inch hard disk drive platter has been machined for this purpose and mounted on a brushless direct current (DC) motor as shown in Fig.~\ref{pic:cho}. The platter is made of aluminium coated with a few thin layers of denser materials, which make the reflected chameleon spectrum similar to the one reflected by the membrane, since aluminium and silicon nitride have similar densities.

\begin{figure}[ht!]
\centering
\includegraphics[width=1\textwidth]{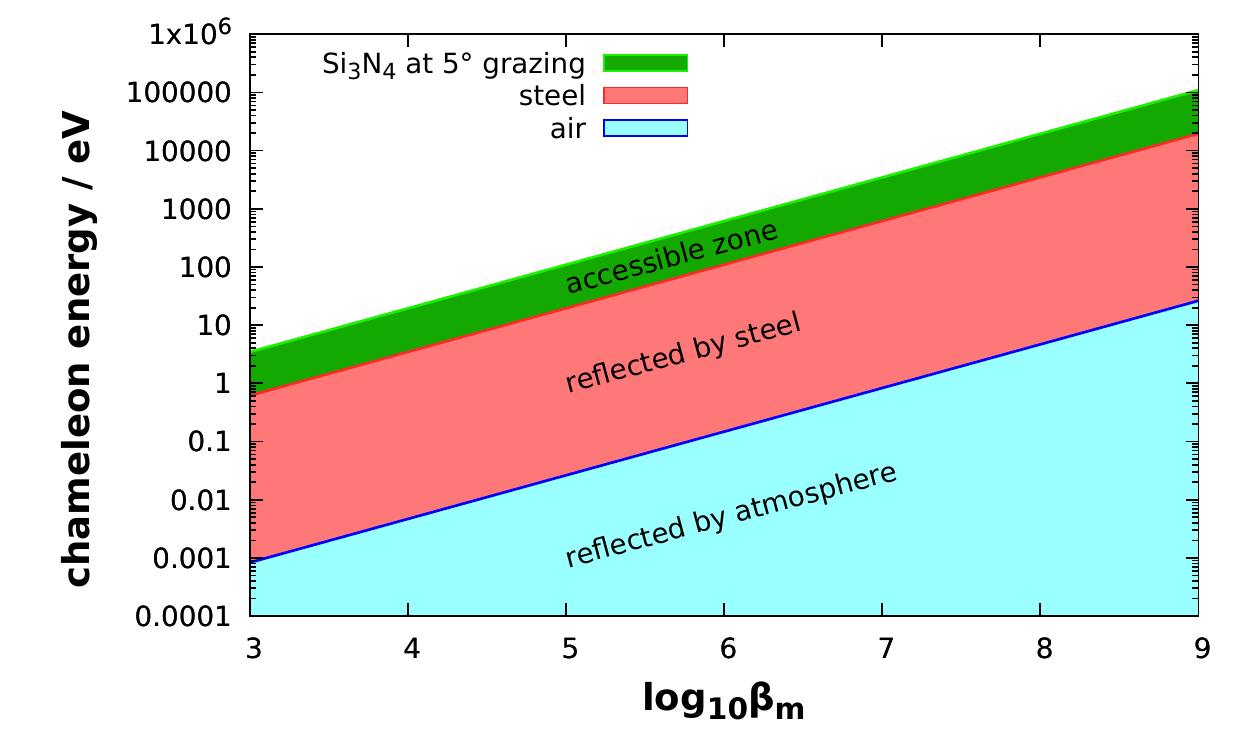}
\caption{The accessible chameleon energy range with the present setup. Only chameleons with masses in the green band between the red and green lines contribute to the radiation pressure on the KWISP sensor.}
\label{pic:azone}
\end{figure}

The chopper disk is mounted at an angle of $5\degree \pm 1\degree$ with respect to the telescope optical axis. This value was chosen as a compromise between the need to maximise the chameleon accessible mass range (see for example Fig.~\ref{pic:azone}), the geometric efficiency of the system and the relative resolution in positioning. The DC motor used is a recycled laser printer polygon mirror motor, modified to rotate a hard disk drive platter. With the platter mounted, the maximum measured rotation speed was about 6000 rpm. Since the disk has 10 roughened segments, this amounts to a maximum modulation frequency of about 1 kHz. The motor is driven by its original electronics, which allows its rotating frequency to be set by a simple pulse frequency modulation. An optical sensor has also been mounted to monitor the exact modulation frequency during all measurements to control fluctuations. Unfortunately the dimensions of the disk constrain the geometric efficiency of the chopper to 10\% calculated as the ratio of the presented chopper segment surface to the chameleon beam surface at the chopper position. 

The modulation frequency has been carefully chosen since it greatly affects the sensitivity. A background noise spectrum is taken to locate a frequency region with low levels of background noise, which is mainly caused by acoustic sources such as vacuum pumps, servo motors, power supplies and others. The environmental noise reaches the membrane inside the vacuum chamber through the chamber walls and mounts. Other noise sources are electronic noise from the power supplies for the laser and the electronics, and stray light reaching the detector. 

\section{KWISP sensor characterisation}

Prior to "on-beam" installation on the CAST magnet, the sensor was characterised in a low background laboratory setting where the influence of  acoustical and mechanical vibrations, which are the main source of noise for the KWISP detector, can be controlled and minimised. In order to determine the sensitivity at possible chopper frequencies, a background spectrum up to 30 kHz was measured. Using laser beam intensities at a maximum with respect to the photo-sensor limits, a displacement sensitivity better than $1\times10^{-15}$m/$\sqrt{\mathrm{Hz}}$ was achieved in the 10's of kHz frequency range.

\begin{figure}[ht!]
\centering
\includegraphics[width=1.0\textwidth]{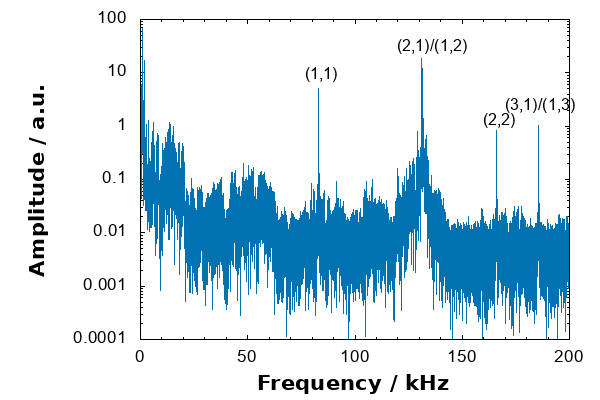}
\caption{Background noise spectrum taken in the laboratory with a the $5 x 5 \  \mathrm{mm^2}$, 100 nm thick membrane. The mechanical vibration peaks can be distinguished from background since their frequencies have well defined ratios (see Table \ref{mempeaks}). Integration time was 9 s with an input beam intensity of 5.0 mW (see also text).}
\label{pic:ms}
\end{figure}

The sensitivity of the system can be easily verified by observing the membrane mechanical vibration modes, as shown in Fig.~\ref{pic:ms}. The amplitude of the mode is a function of the frequency, of the membrane mechanical quality factor, and of the temperature, thus making it easy to estimate. The maximum expected amplitude is lower than $1\times10^{-12}$~m, providing an independent estimate of the device sensitivity.

Since the membrane is square, the following equation links the frequencies of membrane modes:
\begin{equation}
\nu_{j,k}\propto \sqrt{j^2 + k^2} \label{eq:5}
\end{equation}
The peak frequencies seen in the spectrum of Fig.~\ref{pic:ms} are listed in Table \ref{mempeaks}.
The first peak is assumed to be at $j=1 , k=0$. The expected frequencies for the other peaks are then calculated and compared to the measured peaks, as shown in Table \ref{mempeaks}. Matches indicate that some of the peaks in the spectrum are indeed membrane modes.
\begin{table}[ht!]
\centering
\caption{Frequencies of appearance of the peaks observed in the spectrum of Fig.~\ref{pic:ms}. The first column lists the frequencies of the mechanical oscillation peaks, while the second column those expected according to Eq.\ref{eq:5}. Using the first peak as the fundamental vibrational mode, higher mode frequencies can be calculated, and the last two columns give the corresponding mode numbers.}
\label{mempeaks}
\begin{tabular}{|l|l|l|l|}
\hline
Measured frequency / Hz & Calculated frequency / Hz & j & k \\ \hline
82950             & -                    & 1 & 1 \\ \hline
131155            & 131149                & 2  & 1  \\ \hline
165900             & 165941               & 2 & 2 \\ \hline
185482             & 185486               & 3 & 1 \\ \hline
\end{tabular}
\end{table}

\section{Detector calibration}
\begin{figure}[bh!]
\centering
\includegraphics[width=1.0\textwidth]{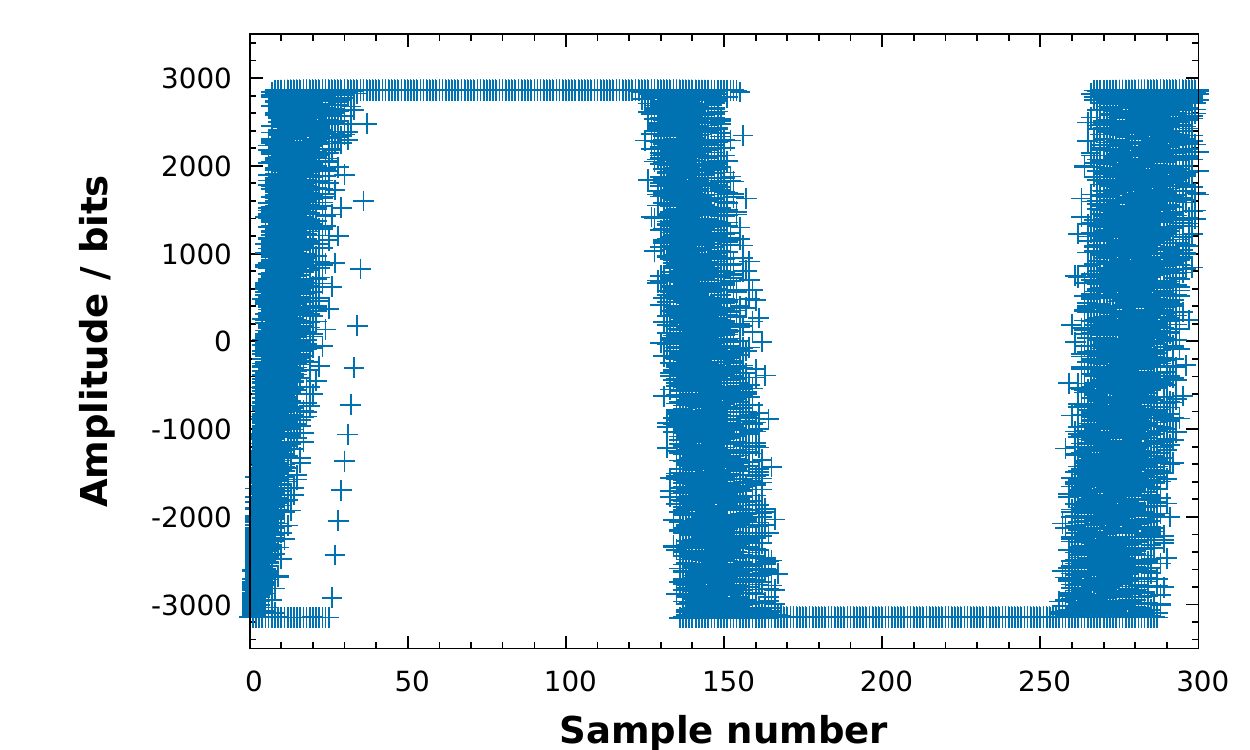}
\caption{Calibration graph. Overlay of the cycles used for calibration. Only the first 532 nm of OPL change are used. }
\label{pic:cg}
\end{figure}

The performance of the detector is checked 
immediately before the start of a data taking run. It is performed by sending a triangular voltage ramp to the feedback piezo and observing changes in the light intensity due to the changes in OPL difference of the Michelson interferometer arms. Typically, the duration of the calibration run is 85 s. It is done with an open feedback loop, which is the only difference with respect to a data taking run. During the calibration, the piezo performs about 850 cycles each approximately $1.2$~$\mu$m  long. The calibration data set collected for one day of data taking is shown in Fig.~\ref{pic:cg}. Two distinct features characterise the plot: Two plateaus, caused by ADC saturation, and the slopes where the interference fringes cross zero. 

\begin{figure}[ht!]
\centering
\includegraphics[width=1.0\textwidth]{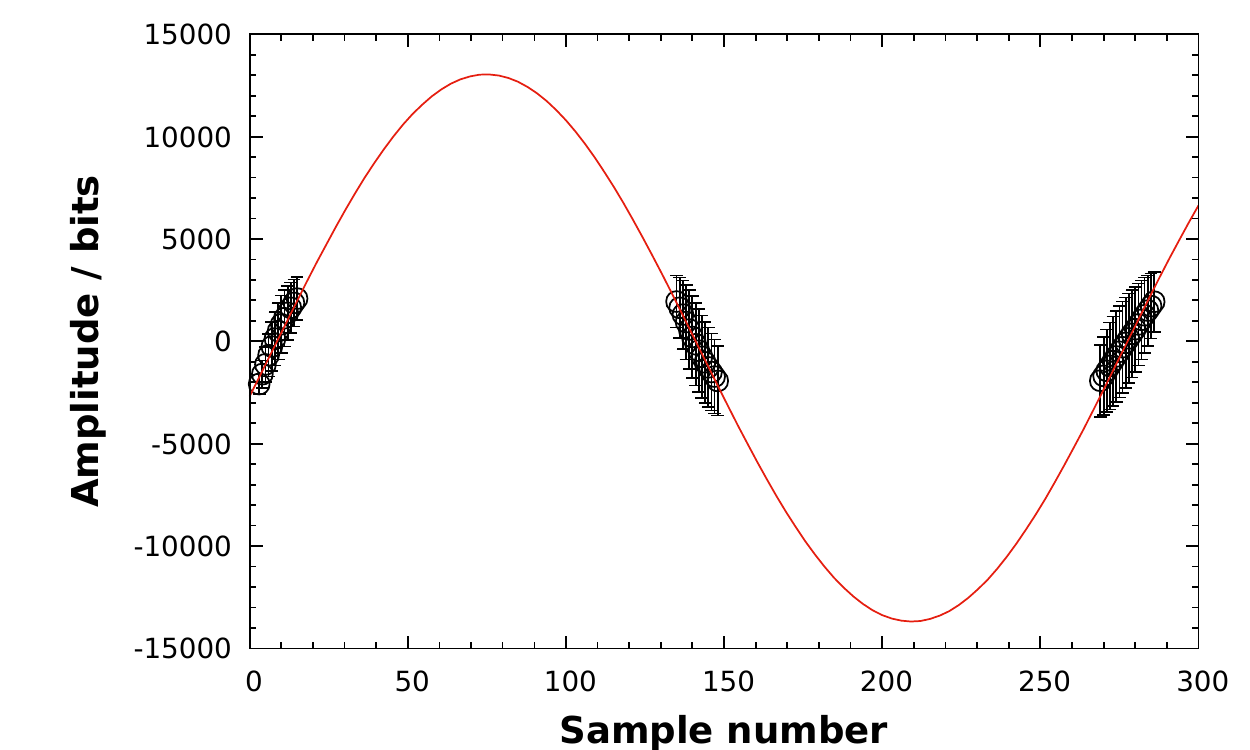}
\caption{Calibration graph showing the result of the sine fit. The data points were fitted with the function $I(x) = A \cdot sin(\omega \cdot x+b)+B$\ and the values $A = (133 \pm 4)\cdot 10^2$, $B = -324 \pm 64$, $\omega = 0.02336 \pm 4$, $\varphi = -0.178 \pm 6$ were derived.}
\label{pic:sine}
\end{figure}

Since the calibration data include also environmental noise, an averaging procedure over different piezo cycles has been applied before fitting the curves. Additionally, noisy data have been rejected by the following procedure:
The derivative of the data for every piezo cycle is calculated and the number of times when the derivative changes sign is extracted. Only data with less than six turning points is kept for the averaging procedure, since larger numbers indicate a change in the OPL difference due to the external noise. From the averaged values an interval of $\pm 2100$~bits around zero is selected in order to exclude the regions with ADC saturation from the calibration data. The remaining data are fitted with the interferometric transfer function $I(x) = A\cdot sin(\omega \cdot x+\varphi)+B$ (see Fig.~\ref{pic:sine}). Here $A$\ is the presumed peak intensity seen by the photodiode, $\omega$\ the circular frequency of a single piezo cycle, $x$\ refers to the sample number, i.e. a time of the measurement, and $\varphi$\ and $B$\ are arbitrary offsets in time and signal amplitude.

The parameters of the fit contain all the information about the detector properties. The most interesting parameter is the derivative of the fit function at its turning point, where the OPL difference is equal to zero. To first approximation this point is also the locking point of the interferometer. 

Considering the number of samples taken for one cycle of the photodiode response function, the laser wavelength $\lambda=532$~nm and the OPL difference $ \delta= (4\pi\, \Delta l)/\lambda$, the calibration constant results $k = A \omega = 3.72\  \mathrm{pm/bit}$. Here we take into account that the change in OPL is twice the displacement of the membrane $\Delta l$.

The error on the calibration constant $k$ is about $3\%$ and is obtained from the error on the parameters of the fit. This value is used in subsequent steps. 

Another value needed to convert the displacement into force is the overall detector response to a known force excitation. This was calibrated in the Laboratory for Quantum and Non-linear Optics at the University of Rijeka. The calibration was done by exciting the membrane with a known periodic force modulated at a frequency near the chopper frequency used during on-beam measurements at CAST. The modulation frequency was chosen in a low background region in order to minimise the time necessary to perform the measurements. An amplitude modulated He-Ne laser beam at $\lambda = 633$~nm was used to excite the membrane and the modulation was achieved by inserting an acousto-optic modulator (AOM) in the light path. This pump beam illuminated the membrane from behind, that is from the opposite side with respect to the 532 nm probe beam. The diameter of the beam, $d = 6$~mm, was estimated by observing the shade of the membrane in the interferometer output, and the average beam power $P = (730\pm~70)$~$\mu$W was measured with a calibrated power meter at the entrance of the vacuum chamber.  The reflectivity of the membrane was separately measured, giving $R = 0.35\pm~0.04$. This results in a peak-to-peak modulation amplitude $A = (510\pm 70)$~$\mu$W or, when expressed as a force, $F = (3.4\pm 0.5)$~pN. Dividing the force by the measured displacement $D = (2.8\pm 0.9)$~fm caused by the external excitation a detector response $K = (1200\pm 400)$~N/m is obtained.

\section{Measurement results}
\begin{figure}[ht!]
\centering
\includegraphics[width=1\textwidth]{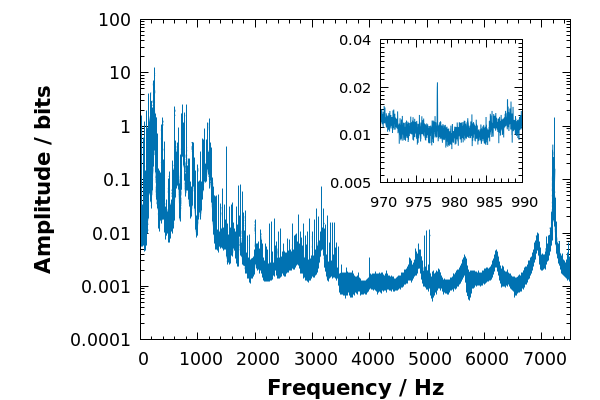}
\caption{Full range single sided amplitude spectrum collected during sun tracking. The acquisition rate was 15258 S/s. The spectrum around the chopper frequency 984 Hz is shown in the inset.}
\label{pic:f5}
\end{figure}

Following online detector calibration, the CAST telescope is set in tracking mode and the magnet movement begins. After approximately half an hour of movement the telescope starts actually tracking the Sun. The tracking lasts for about 90 minutes and the data from the detector are acquired and transferred to the control room. For practical reasons, data are stored in files containing 15 minutes of data taking, resulting in six data files per solar tracking. The data taking campaign of this test run lasted for ten days. The focus was on detector test and commissioning. The outcome from a single solar tracking is presented in this work to illustrate the potential of this method. The acquired data from each data file are separately analysed in the offline analysis where the Fourier transforms of different data sets are averaged to obtain the final spectrum. The chopper frequency was set to 984 Hz, therefore, a chameleon signal would be expected to appear only in a few bins around the preset frequency. This particular frequency was chosen considering the low noise floor in the spectrum in this region as can be seen in Fig.~\ref{pic:f5}.
During the measurements, intense low frequency background noise was observed around 250 Hz, which at greater sensitivity would dominate the measurements due to the finite dynamic range of the DAQ.

The final result can be seen in the inset of the figure and the obtained noise limit, calculated as an average value in a band of $\pm$~1~Hz around the chopper frequency, is $(1.0\,\pm\,0.2)\cdot 10^{-2}$ bits, giving a bound on membrane displacement at the frequency of 984 Hz of $d_{lim}$ = 37.2~fm $\pm$ 3\% (calibration) $\pm$~ 20\% (background).

\begin{figure}[ht!]
\centering
\includegraphics[width=1\textwidth]{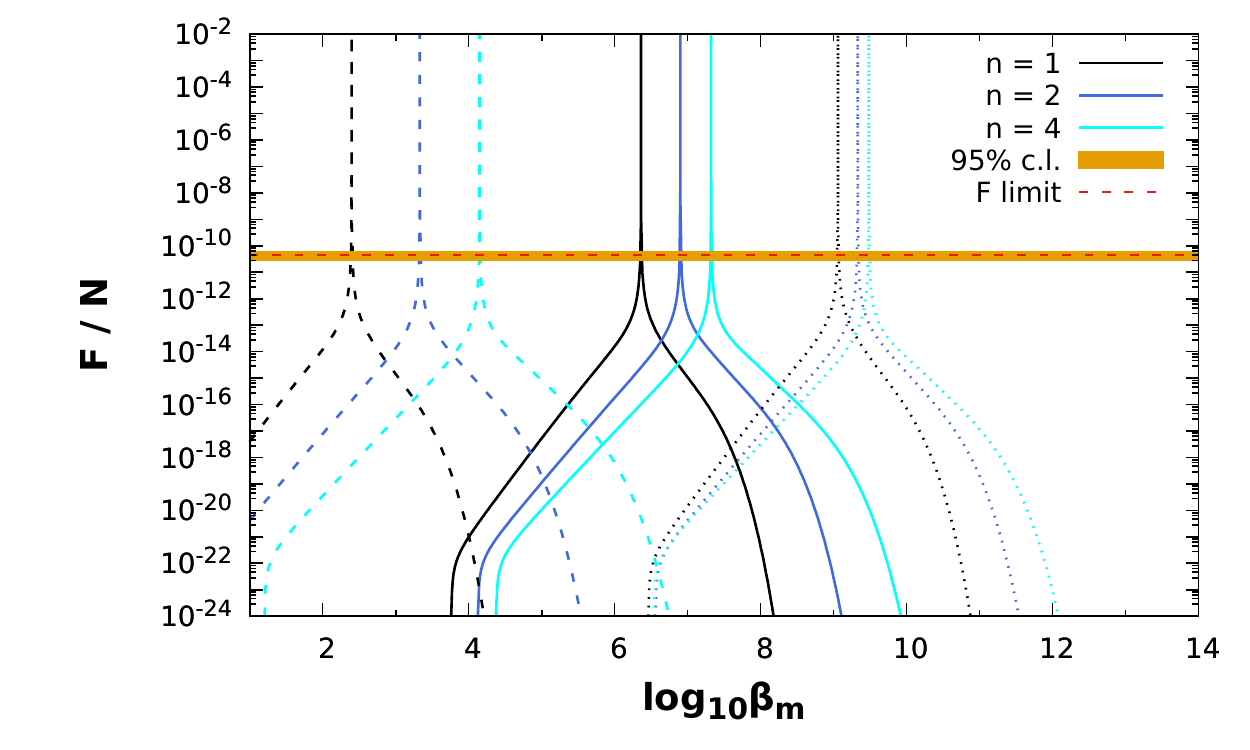}
\caption{Expected force at the sensor calculated from the solar chameleon flux assuming $\beta_\gamma = 10^{10.32}$, interaction zone length $L_i = 0.05 \cdot R_{Sun}$ and detector parameters. Solid lines correspond to a dark energy scale $\Lambda = 2.4 \cdot 10^{-3}$~eV, dashed ones to $\Lambda = 1 \cdot 10^{-5}$~eV, and dotted lines to $\Lambda = 0.1$~eV. The interaction zone length is an upper limit according to current Solar models. The flux and consequently the expected force are scaled accordingly for values down to $L_{i} = 0.01 \cdot R_{Sun}$ while the resonance peak is arbitrarily limited to $F = 10^{-2}$~N by the axis scale. The orange band represents 95\% confidence interval centred at the average force value shown by the red dashed line.}
\label{pic:chforce}
\end{figure}

Using the measured detector response the limit on the sensed force becomes $F_{lim} = 44\pm 18$~pN where the uncertainties from the detector response and observed background have been added in quadrature. This value is then compared with
 the expected force on the membrane calculated from the solar chameleon spectrum $\Phi_{cham}(\omega)$. The calculation proceeds by multiplying the chameleon spectrum by energy and integrating it between limits given by the chameleon effective masses in steel and silicon nitride at grazing incidence. This gives the momentum carried in the given energy range for each $\beta_m$. Additional factors have been taken into account, such as the distance from Sun to Earth, the 10\% geometrical chopping efficiency, and the 43 mm magnet bore diameter. Only the component of the chameleon momentum perpendicular to the membrane surface has been considered. Note that the CAST magnet bore diameter limits the surface of the telescope seen by the chameleons.
 Immediately prior to hitting the membrane the chameleons are propagating in vacuum and their effective mass can be assumed as $m \approx 0$. In the KWISP detector all relevant distances to vacuum chamber walls are sufficiently large for the effective potential to reach its minimum and chameleon velocities remain relativistic. Finally, the expected force acting on the membrane can be calculated using all the additional factors and compared to the measured value in Fig.~\ref{pic:chforce}.
 
\section{Results}
\begin{figure}[ht!]
\centering
\includegraphics[width=1\textwidth]{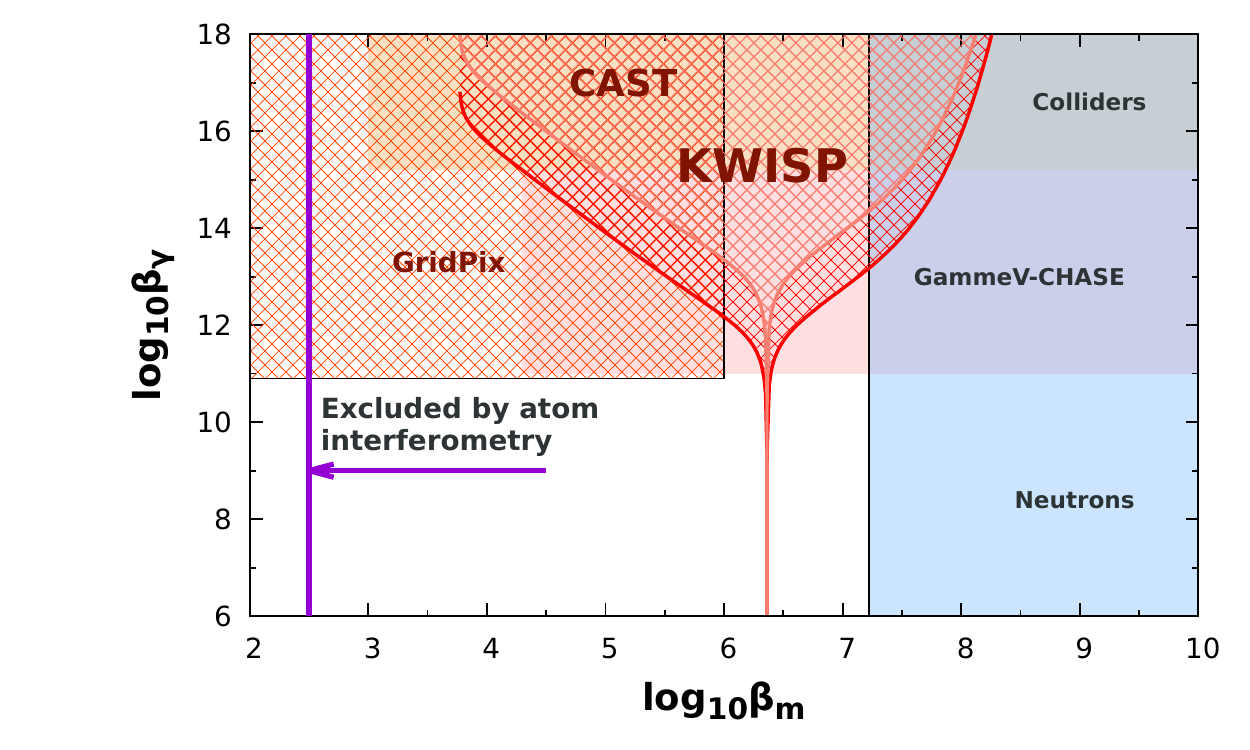}
\caption{The exclusion limit for Chameleons at CAST are shown with hatched pattern. The KWISP measurement of this work is illustrated in red colour and comparison to other experimental techniques (\cite{gammev,gammevchase,collider,atomin}) is made, assuming a dark energy scale $\Lambda = 2.4 \cdot 10^{-3}$ eV and $n = 1 $. The limit from the previous CAST measurement \cite{castchin} is based on chameleon-to-photon conversion only. In the parameter space of realistic solar models the expected chameleon production maximum, in darker shade, takes place for $B=30$~T and the tachocline length $L_i = 0.05 R_{Sun}$ while the minimum, in lighter shade, is expected for $B=10$~T and the tachocline length $L_i = 0.01 R_{Sun}$. }
\label{pic:chexplot}
\end{figure}

 The solar chameleon model predicts a mass-dependent force which is displayed in Fig.~\ref{pic:chforce} for a fixed $\beta_\gamma = 10^{10.32}$. The calculated solar flux (Eq. \ref{eq:flux}) depends on one parameter only, the coupling constant of chameleons to photons $\beta_{\gamma}$. The coupling to matter $\beta_{m}$ enters the calculation only through the effective mass of chameleons in the Sun, and introduces a cutoff in the spectrum. Otherwise, once the chameleons leave the Sun and enter a vacuum region, their mass becomes zero and the related energy is added to their momentum, thus the chameleon spectrum exhibits the same energy dependence as the original photon spectrum.  
 This allows us to consider the measured force proportional to the square of the $\beta_{\gamma}$, or written in another way the relation becomes
 \begin{equation}
\beta_{\gamma}=\sqrt{\frac{F}{A}}.     
 \end{equation}
Here the parameter $A$ absorbs all variables in Eq. \ref{eq:flux} except for the coupling constant to photons. We observed no chameleon signal above background in our set up; the obtained force limit is $(44 \pm 18)$~pN. In Fig.~\ref{pic:chexplot} we showthe corresponding exclusion limit in the $\beta_m$ -- $\beta_\gamma$ parameter plane for fixed values of the remaining chameleon parameters, $\Lambda=2.4~$meV and $n=$1.

Since no chameleon signal above background is observed in our set-up at a force limit of $(44 \pm 18)$~pN the calculated exclusion limits are presented in Fig.~\ref{pic:chexplot} in the $\beta_m$ and $\beta_\gamma$ parameter plane assuming $\Lambda=2.4~$meV and $n=$1.

\section{Conclusions}
The displacement sensitivity of the KWISP detector was calibrated in an optics laboratory while the detector response to a known radiation pressure force, provided by photon beam, was also measured in order to provide the conversion parameter from displacement to force. 
The results of the Solar chameleon measurements at CAST taken during July 2017 put the limit on the force acting at the membrane at $(44 \pm 18)$~pN. Using this result, combined with the expected chameleon flux at the detector, allows one to define an exclusion region in the $\beta_m-\beta_\gamma$ plane as shown in Fig.~\ref{pic:chexplot}. 
This region is superimposed on areas explored by other experiments (\cite{gammev,gammevchase,collider}) including atom interferometry \cite{atomin,atomith,atomith2017},  which claims to exclude all $\beta_m \geq 10^{2}$, however based on purely virtual chameleon exchange, and "afterglow" experiments, which rely on two successive photon-chameleon and chameleon-photon conversions. In the previous CAST measurement \cite{castchin} the first photon-to-chameleon conversion happens in the sun, and the second reverse conversion in the magnet provides the photons to be detected, while the measurement reported here with the KWISP detector is directly sensitive to real chameleon interactions, and in this sense it is complementary to the previously existing exclusion limits.

The sensitivity of the KWISP opto-mechanical sensor can be improved by at least two orders of magnitude with already ongoing upgrades. This, combined with a reasonable integration time, will open previously uncharted territory in the $\beta_m$ $\beta_\gamma$ plane for real chameleon interactions.

\section{Acknowledgments}
We thank CERN for hosting the experiment and for technical support to operate the magnet and cryogenics.
We acknowledge support from NSERC (Canada), MSE (Croatia), University of Rijeka with grant no 13.12.2.2.09, 18-126, EU fund projects KK.01.1.1.01.0001 and RC.2.2.06.-0001, BMBF under the grant numbers 05 CC2EEA/9 and 05 CC1RD1/0, DFG under grant numbers HO 1400/7-(1-3) and EXC-153, the University of Freiburg (Germany), GSRT (Greece), NSRF: Heracleitus II, RFFR (Russia), the Spanish Ministry of Economy and Competitiveness (MINECO) under Grants No. FPA2011-24058 and No. FPA2013-41085-P (grants partially funded by the European Regional Development Fund, ERDF/FEDER), the European Research Council (ERC) under grant ERC-2009-StG-240054 (T-REX), the Turkish Atomic Energy Authority (TAEK) and IBS (Korea) with code IBS-R017-D1-2017-a00. Part of this work was performed under the auspices of the U.S. Department of Energy by Lawrence Livermore National Laboratory under Contract No. DE-AC52-07NA27344.

\bibliography{refs}

\begin{thebibliography}{10}
\expandafter\ifx\csname url\endcsname\relax
  \def\url#1{\texttt{#1}}\fi
\expandafter\ifx\csname urlprefix\endcsname\relax\def\urlprefix{URL }\fi
\expandafter\ifx\csname href\endcsname\relax
  \def\href#1#2{#2} \def\path#1{#1}\fi

\bibitem{art5}
J.~Khoury, A.~Weltman, Chameleon cosmology, Phys. Rev. D 69 (2004) 044026.
\newblock \href {http://dx.doi.org/doi.org/10.1103/PhysRevD.69.044026}
  {\path{doi:doi.org/10.1103/PhysRevD.69.044026}}.

\bibitem{Sikivie:1983jw}
P.~Sikivie, {Experimental Tests of the "Invisible" Axion}, Physical Review
  Letters 51~(16) (1983) 1415--1417.
\newblock \href {http://dx.doi.org/doi.org/10.1103/PhysRevLett.51.1415}
  {\path{doi:doi.org/10.1103/PhysRevLett.51.1415}}.

\bibitem{fifthforce}
A.~Upadhye, S.~S. Gubser, J.~Khoury, Unveiling chameleon fields in tests of the
  gravitational inverse-square law, Phys. Rev. D 74 (2006) 104024.
\newblock \href {http://dx.doi.org/doi.org/10.1103/PhysRevD.74.104024}
  {\path{doi:doi.org/10.1103/PhysRevD.74.104024}}.

\bibitem{fifthforce1}
A.~Upadhye, Dark energy fifth forces in torsion pendulum experiments, Phys.
  Rev. D 86 (2012) 102003.
\newblock \href {http://dx.doi.org/doi.org/10.1103/PhysRevD.86.102003}
  {\path{doi:doi.org/10.1103/PhysRevD.86.102003}}.

\bibitem{baum}
S.~Baum, G.~Cantatore, D.~Hoffmann, M.~Karuza, Y.~Semertzidis, A.~Upadhye,
  K.~Zioutas, Detecting solar chameleons through radiation pressure, Physics
  Letters B 739 (2014) 167 -- 173.
\newblock \href {http://dx.doi.org/doi.org/10.1016/j.physletb.2014.10.055}
  {\path{doi:doi.org/10.1016/j.physletb.2014.10.055}}.

\bibitem{reflection}
P.~Brax, C.~van~de Bruck, A.-C. Davis, D.~F. Mota, D.~Shaw, Testing chameleon
  theories with light propagating through a magnetic field, Phys. Rev. D 76
  (2007) 085010.
\newblock \href {http://dx.doi.org/doi.org/10.1103/PhysRevD.76.085010}
  {\path{doi:doi.org/10.1103/PhysRevD.76.085010}}.

\bibitem{art3}
K.~{Baker}, A.~{Lindner}, A.~{Upadhye}, K.~{Zioutas}, A chameleon helioscope,
  \,\,\href {http://arxiv.org/abs/1201.0079} {\path{arXiv:1201.0079}}.

\bibitem{art4}
M.~{Karuza}, G.~{Cantatore}, A.~{Gardikiotis}, D.~H.~H. {Hoffmann}, Y.~K.
  {Semertzidis}, K.~{Zioutas}, \mbox{KWISP}: An ultra-sensitive force sensor
  for the dark energy sector, Physics of the Dark Universe 12 (2016) 100--104.
\newblock \href {http://arxiv.org/abs/1509.04499} {\path{arXiv:1509.04499}},
  \href {http://dx.doi.org/doi.org/10.1016/j.dark.2016.02.004}
  {\path{doi:doi.org/10.1016/j.dark.2016.02.004}}.

\bibitem{art7}
M.~Kuster, et~al., The x-ray telescope of \mbox{CAST}, New J. Phys. 9 (2007)
  169--173.
\newblock \href {http://dx.doi.org/doi.org/10.1088/1367-2630/9/6/169}
  {\path{doi:doi.org/10.1088/1367-2630/9/6/169}}.

\bibitem{homo}
P.~Piergentili, L.~Catalini, M.~Bawaj, S.~Zippilli, N.~Malossi, R.~Natali,
  D.~Vitali, G.~D. Giuseppe, Two-membrane cavity optomechanics, New Journal of
  Physics 20~(8) (2018) 083024.

\bibitem{bachor2004guide}
H.~Bachor, T.~Ralph, A Guide to Experiments in Quantum Optics, Wiley, 2004.

\bibitem{RPTY}
\href{http://redpitaya.com/}{Red {P}itaya {V}ersion 1.1}.
\newline\urlprefix\url{http://redpitaya.com/}

\bibitem{brax}
P.~Brax, A.~Lindner, K.~Zioutas, Detection prospects for solar and terrestrial
  chameleons, Phys. Rev. D 85 (2012) 043014.
\newblock \href {http://dx.doi.org/doi.org/10.1103/PhysRevD.85.043014}
  {\path{doi:doi.org/10.1103/PhysRevD.85.043014}}.

\bibitem{solar_cham}
P.~Brax, K.~Zioutas, Solar chameleons, Phys. Rev. D 82 (2010) 043007.
\newblock \href {http://dx.doi.org/doi.org/10.1103/PhysRevD.82.043007}
  {\path{doi:doi.org/10.1103/PhysRevD.82.043007}}.

\bibitem{ppath}
R.~Mitalas, K.~R. Sills, On the photon diffusion time scale for the sun, The
  Astrophysical Journal 401 (1992) 759.
\newblock \href {http://dx.doi.org/10.1086/172103} {\path{doi:10.1086/172103}}.

\bibitem{vinyoles}
N.~Vinyoles, et~al., A new generation of standard solar models, The
  Astrophysical Journal 835 (2017) 202.
\newblock \href {http://dx.doi.org/doi.org/10.3847/1538-4357/835/2/202}
  {\path{doi:doi.org/10.3847/1538-4357/835/2/202}}.

\bibitem{castchin}
V.~{Anastassopoulos}, {\textit{et al.}}~{[CAST Collaboration]}, Improved search
  for solar chameleons with a {GridPix} detector at {CAST}, JCAP 01 (2019) 032.
\newblock \href {http://dx.doi.org/doi.org/10.1088/1475-7516/2019/01/032}
  {\path{doi:doi.org/10.1088/1475-7516/2019/01/032}}.

\bibitem{ANASTASSOPOULOS2015172}
V.~{Anastassopoulos}, {\textit{et al.}}~{[CAST Collaboration]}, Search for
  chameleons with {CAST}, Physics Letters B 749 (2015) 172 -- 180.
\newblock \href {http://dx.doi.org/doi.org/10.1016/j.physletb.2015.07.049}
  {\path{doi:doi.org/10.1016/j.physletb.2015.07.049}}.

\bibitem{gammev}
A.~Upadhye, J.~H. Steffen, A.~Weltman, Constraining chameleon field theories
  using the {GammeV} afterglow experiments, Phys. Rev. D 81 (2010) 015013.
\newblock \href {http://dx.doi.org/doi.org/10.1103/PhysRevD.81.015013}
  {\path{doi:doi.org/10.1103/PhysRevD.81.015013}}.

\bibitem{gammevchase}
J.~H. Steffen, et~al., Laboratory constraints on chameleon dark energy and
  power-law fields, Phys. Rev. Lett. 105 (2010) 261803.
\newblock \href {http://dx.doi.org/doi.org/10.1103/PhysRevLett.105.261803}
  {\path{doi:doi.org/10.1103/PhysRevLett.105.261803}}.

\bibitem{collider}
P.~Brax, C.~Burrage, A.-C. Davis, D.~Seery, A.~Weltman,
  \href{https://doi.org/10.1088%2F1126-6708%2F2009%2F09%2F128}{Collider
  constraints on interactions of dark energy with the standard model}, Journal
  of High Energy Physics 2009~(09) (2009) 128--128.
\newblock \href {http://dx.doi.org/10.1088/1126-6708/2009/09/128}
  {\path{doi:10.1088/1126-6708/2009/09/128}}.
\newline\urlprefix\url{https://doi.org/10.1088%2F1126-6708%2F2009%2F09%2F128}

\bibitem{atomin}
P.~Hamilton, M.~Jaffe, P.~Haslinger, Q.~Simmons, H.~M{\"u}ller, J.~Khoury,
  \href{https://science.sciencemag.org/content/349/6250/849}{Atom-interferometry
  constraints on dark energy}, Science 349~(6250) (2015) 849--851.
\newblock \href
  {http://arxiv.org/abs/https://science.sciencemag.org/content/349/6250/849.full.pdf}
  {\path{arXiv:https://science.sciencemag.org/content/349/6250/849.full.pdf}},
  \href {http://dx.doi.org/10.1126/science.aaa8883}
  {\path{doi:10.1126/science.aaa8883}}.
\newline\urlprefix\url{https://science.sciencemag.org/content/349/6250/849}

\bibitem{atomith}
B.~Elder, et~al., Chameleon dark energy and atom interferometry, Phys. Rev. D
  94 (2016) 044051.
\newblock \href {http://dx.doi.org/doi.org/10.1103/PhysRevD.94.044051}
  {\path{doi:doi.org/10.1103/PhysRevD.94.044051}}.

\bibitem{atomith2017}
M.~Jaffe, et~al., Testing sub-gravitational forces on atoms from a miniature
  in-vacuum source mass, Nature Physics 13 (2017) 938.
\newblock \href {http://dx.doi.org/doi.org/10.1038/nphys4189}
  {\path{doi:doi.org/10.1038/nphys4189}}.

\end{thebibliography}

\end{document}